\documentclass[preprint1]{aastex62}

\shortauthors{Miao et al.}
\usepackage{url}
\usepackage{hyperref}

\newcommand{\speed}[1]{#1 km~s${}^{-1}$}
\newcommand{\acc}[1]{#1 m~s${}^{-2}$}

\newcommand{\degree}{\ensuremath{^\circ}}

\begin{document}

\title{A Quasi-periodic propagating wave and EUV waves excited simultaneously in a solar eruption event}


\correspondingauthor{Y. H. Miao}
\email{myh@ynao.ac.cn}
\author[0000-0003-2183-2095]{Y. H. Miao}
\affiliation{Yunnan Observatories, Chinese Academy of Sciences, Kunming, 650216, China}
\affiliation{Key Laboratory of Geospace Environment, Chinese Academy of Sciences, University of Science $\&$ Technology of China, Hefei 230026, China}
\affiliation{Radio Cosmology Lab, Department of Physics, Faculty of Science, University of Malaya, 50603 Kuala Lumpur, Malaysia.}
\affiliation{Department of Physics and Astronomy, King Saud University,
PO Box 2455, Riyadh 11451, Saudi Arabia}
\affiliation{University of Chinese Academy of Sciences, Beijing 100049, China}

\author{Y. Liu}
\affiliation{Yunnan Observatories, Chinese Academy of Sciences, Kunming, 650216, China}
\affiliation{Key Laboratory of Geospace Environment, Chinese Academy of Sciences, University of Science $\&$ Technology of China, Hefei 230026, China}
\affiliation{Radio Cosmology Lab, Department of Physics, Faculty of Science, University of Malaya, 50603 Kuala Lumpur, Malaysia.}
\affiliation{Department of Physics and Astronomy, King Saud University,
PO Box 2455, Riyadh 11451, Saudi Arabia}

\author{Y. D. Shen}
\affiliation{Yunnan Observatories, Chinese Academy of Sciences, Kunming, 650216, China}
\affiliation{Key Laboratory of Geospace Environment, Chinese Academy of Sciences, University of Science $\&$ Technology of China, Hefei 230026, China}

\author{H. B. Li}
\affiliation{Yunnan Observatories, Chinese Academy of Sciences, Kunming, 650216, China}
\affiliation{University of Chinese Academy of Sciences, Beijing 100049, China}

\author{Z. Z. Abidin}
\affiliation{Radio Cosmology Lab, Department of Physics, Faculty of Science,
University of Malaya, 50603 Kuala Lumpur, Malaysia.}

\author{A. ELMHAMDI}
\affiliation{Department of Physics and Astronomy, King Saud University,
PO Box 2455, Riyadh 11451, Saudi Arabia}

\author{A. S. KORDI}
\affiliation{Department of Physics and Astronomy, King Saud University,
PO Box 2455, Riyadh 11451, Saudi Arabia}

\begin{abstract}


Quasi-periodic fast-propagating (QFP) magnetosonic waves and extreme
ultraviolet (EUV) waves were proposed to be driven by solar flares and coronal
mass ejections (CMEs), respectively. In this Letter, we present a detailed
analysis of an interesting event in which we find that both QFP magnetosonic
waves and EUV waves are excited simultaneously in one solar eruption event. The
co-existence of the two wave phenomena offers an excellent opportunity to
explore their driving mechanisms. The QFP waves propagate in a funnel-like loop
system with a speed of 682--837 \speed{} and a lifetime of 2 minutes. On the
contrary, the EUV waves, which present a faster component and a slower
component, propagate in a wide angular extent, experiencing reflection and
refraction across a magnetic quasi-separatrix layer. The faster component of
the EUV waves travels with a speed of 412--1287 \speed{}, whereas the slower
component travels with a speed of 246--390 \speed{}. The lifetime of the EUV
waves is $\sim$15 minutes. It is revealed that the faster component of the EUV
waves is cospatial with the first wavefront of the QFP wave train. Besides,
The QFP waves have a period of about 45$\pm$5 seconds, which is absent in the
associated flares. All these results imply that QFP waves can also be excited
by mass ejections, including CMEs or jets.

\end{abstract}

\section{Introduction}
Waves are ubiquitous in the solar atmosphere, and have been investigated
extensively. In the past several decades, various types of waves have been
observed in the solar atmosphere, such as Moreton waves \citep{moreton1960,
shen12b,krause2018}, coronal extreme-ultraviolet (EUV) waves
\citep[e.g.,][]{thompson98, thompson99, liuw10, yang13, liuw2014, muhr14,
shen18d}, slow \citep[e.g.,][]{nakariakov2011} and fast \citep{ofman02, liuw11,
liuw12, yuanding2013, zhangyuzong2015, ofman2018,miao2018} modes magnetosonic waves.
Besides these traveling waves, there exist various kinds of stationary waves
trapped in coronal loops and filaments, which lead to coronal loop and filament
oscillations \citep{nakariakov2001, nakariakov2005, chenpf2008, liuw12,
lit2012b, shen14a, shen14b, zhou2018}. Waves and oscillatory phenomena in the
solar atmosphere have been attractive since they are considered to be very
important for coronal heating or be used to diagnose the magnetic field in
which the waves are propagating \citep{nakariakov1999a, nakariakov1999b,
nakariakov2001,doorsselaere2008, chenpf2009a, shen12,shen13,ofman2018,shen18d}.

Large-scale extreme ultraviolet (EUV) waves were first detected by the
Extreme-ultraviolet Imaging Telescope \citep[EIT;][]{dela95}) aboard the
{\sl Solar Heliospheric Observatory}, and was originally called ``EIT waves"
\citep{thompson98, thompson99}. In the past two decades, a large number of
studies have been conducted to explore the physical nature and driving
mechanism of this wave phenomenon. Regarding its physical nature, early
researchers considered that ``EIT waves" are probably the coronal counterparts
of the chromospheric Moreton waves , i.e., they are fast-mode waves or shock
waves \citep{thompson98, thompson99, wang00,wu01,ofman02, schmidt10, shen12a,
shen17b, shen18a, shen18d}. On the other hand, several non-wave models were
suggested \citep{dela00,2008SoPh..247..123D, liuw10, liuw12}. In particular,
\citet{chenpf2002, chenpf2005} proposed that there should be two components of
EUV waves associated with a coronal mass ejection (CME) event, i.e., a
fast-modea wave (or shock wave) and a slowly moving apparent wave. Regarding
the driving mechanism, some authors proposed that EUV waves are excited by the
pressure pulse inside the flare \citep[e.g.,][]{khan2002, hudson2003,
warmuth2004}, while others considered that they are driven by CMEs or jets
\citep[e.g.,][]{cliver1999, chenpf2002, chenpf2009, lit2012a, shen12a,
shen17a}. It is now generally believed that the large-scale EUV waves, either
the faster component or the slower one, are driven by mass ejections, either
CMEs or jets \citep[see][for reviews]{warmuth2015, chenpf2016}.

In recent years, another type of wave phenomenon, i.e., quasi-periodic
fast-propagating (QFP) magnetosonic waves, became a hot topic, which was first
reported by \citet{liuw12} using high spatial and temporal resolution
observations taken by the Atmospheric Imaging Assembly (AIA; \citealt{lemen12})
on board the {\em Solar Dynamics Observatory} ({\em SDO}; \citealt{pesnell12}).
The QFP waves are observed as a wave train with multiple arc-shaped fronts,
which propagate along funnel-like coronal loops. According to \citet{liuw2014},
the period, speed, and deceleration of QFP waves are in ranges of 25--400
seconds, \speed{500--2200}, and \acc{1--4}, respectively. In the past seven
years, many authors claimed that QFP waves have an intimate relationship with
the accompanying flare and their initial positions are often seen to be in a
fewa megameters from the flare kernel \citep{liuw11, liuw12, shen12b,
yuanding2013, kumar2017, shen18a, shen18b}. \citet{ofman2018} presented
numerical simulations in order to understand the physics of QFP waves. Their
simulation results tend to support that the pressure pulse inside a flare can
generate QFP waves.

In this letter, we analyze a QFP wave and EUV waves that appeared
simultaneously in the flare/CME event on 2011 March 10, which was captured by
the {\sl SDO}/AIA and the {\sl SDO} Helioseismic and Magnetic Imager (HMI;
\citealt{schoul12}). Among the 10 wavelengths of the {\sl SDO}/AIA, only the
171 \AA\ (\ion{Fe}{9}; $\lg${\em T}=5.8) channel clearly showed the QFP wave,
whereas the EUV waves can be seen in various EUV channels, in particular the
193 \AA\ channel (\ion{Fe}{12}; $\lg${\em T}=6.1). We concentrate on the 171
\AA\ and 193 \AA\ channels in this study. Analysis results are introduced in
Section \ref{res}, and conclusions and discussions are presented in Section
\ref{dis}. The co-existence of the two wave phenomena offers an excellent
opportunity to explore their driving mechanisms.

\section{Results}\label{res}

A {\sl GOES} C4.0 flare, whose start, peak, and end times were at about
06:40:00, 06:41:12, and 06:53:12 UT respectively, occurred in NOAA active
region AR11166 on 2011 March 10. The CME was observed by {\sl STEREO}
spacecraft in two viewpoints\footnote{\url{https://cdaw.gsfc.nasa.gov/movie/make_javamovie.php?img1=stb_co1s&img2=sta_co1s&stime=20110310_0400&etime=20110310_0800}}
since about 06:50 UT. The morphology of the CME is similar to that in
\citet{patsourakos09} and will be investigated in a separate paper.
Here, we focus on the driving mechanisms of the QFP wave and the EUV waves.
Note that this EUV wave event was not listed in the EUV wave catalog compiled
by \citet{nitt13}.

\begin{figure}
\epsscale{1.0} \plotone{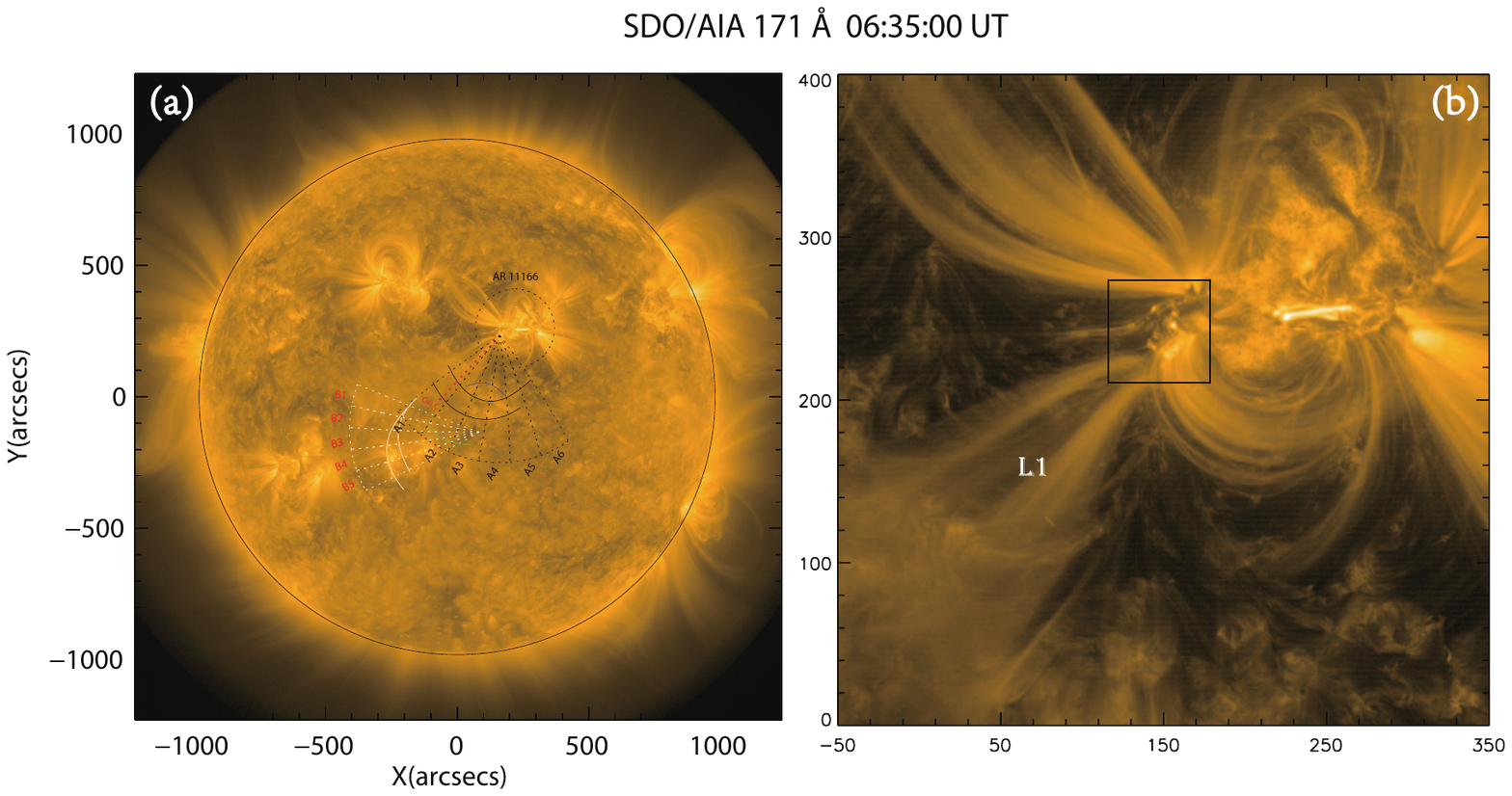}
\caption{Panel (a): {\sl SDO}/AIA 171 \AA\ full-disk image shows six {15\degree} (A1-A6), five {10\degree} (B1-B5)
wide sectors. The dotted curve (Cut C) presents the path used to obtain time-distance diagram. The black and blue lines
represent the primary EUV wave and reflection, respectively. The green and and white lines show the two
refracted waves, respectively. Panel (b): the L1 represents the funnel-like loop system. The box displays the flare active region.
\label{slice_position}}
\end{figure}

At about 06:40:38 UT, the first arc-shaped wavefront of the QFP wave was
seen to propagate along the funnel-like loop system at the southeastern
periphery of AR11166. The wavefront was visible for 2 minutes. Panel (a) of
Figure \ref{slice_position} shows the overview of the QFP and EUV wave
propagation coverage on the AIA 171 \AA\ full-disk image, where several slices
are chosen for plotting time-distance diagrams. We use a semi-automatic
method to get time-distance diagrams from {15\degree} wide sectors along
A1--A6 on the solar surface \citep{podladchikova05, liuw10, long11, lit2012a}.
The same method is applied to sectors B1--B5, which are {10\degree} wide.
The red dotted curve (Cut C) shows the path of the QFP wave train, whereas the
black, blue, and white lines represent the initial EUV waves, their reflection,
and refraction, respectively. Panel (b) of Figure \ref{slice_position} shows
the funnel-like loop system (L1) and the active region (see the black box). One
can see a cluster of funnel-like loops rooted in AR11166. The QFP wave train is
observed to propagate along L1 in the 171 \AA\ images. It should be noted that
the QFP wave could be observed only in the 171 \AA\ channel along the
funnel-like loop system (see L1 in panel(b) of Figure \ref{slice_position}).
With the {\em SDO}/AIA and {\em STEREO}/Ahead EUVI data and using the
scc$\_$measure.pro code in the standard Solar Software (SSW), we also measure
the height of the funnel-like loop system, which is about 40 Mm.

The QFP and EUV waves morphologies and evolutions are displayed in Figure
\ref{wave}. Panels (a1)-(a6) and (b1)-(b6) are running difference images (i.e.,
from each image we subtract the one 12 seconds earlier) at 171 \AA\ and 193
\AA, respectively. At about 06:40:38 UT, the QFP wave was launched in the
southeast of the AR11166 and its initial appearance position was about 36 Mm to
the flare kernel. The start time of the QFP wave (06:40:38 UT) was slightly
lagging behind the beginning of the accompanying flare (06:40:00 UT). Panel
(a1) of Figure \ref{wave} shows the profiles of the wavefronts of the QFP wave,
which are indicated by the green lines. In panel (b1) of Figure \ref{wave}, the
EUV wavefront can be seen clearly at 06:41:22 UT. The EUV wave front can also
be seen in Figure \ref{wave}(a2), as indicated by the shorter green line. We
present an animation (animation.mpeg) to display the detailed evolution of the
QFP and EUV waves in the online journal. Combing panels (a2) and (b1) of Figure
\ref{wave}, we find that the EUV wavefront is in juxtaposition with the first
wavefront of the QFP wave train, which might indicate that the EUV wave is
probably the counterpart of the first wavefront of the QFP wave train.

In panel (a3) of Figure \ref{wave}, the relevant coronal structures are
presented. The upper white arrow indicates the L1 loop system. The two red
arrows indicate a magnetic separatrix surface \citep{thomp09,liuw12,schmidt10}.
When the EUV wavefront approaches the magnetic separatrix surface, the
wavefront became weaker and weaker, and then disappeared in the structure.
About two minutes later (at about 06:45:12 UT), a new and small wavefront (see
the green line in panels (a3)--(a4) of Figure \ref{wave}) appeared to the
southeast of the separatrix structure. Since there is no other erupting source
around the magnetic separatrix structure, we suppose that the new wavefront is
probably a refracted wave produced during the wave passage. In the meantime, a
reflected wave is also observed to the north of the separatrix structure, which
is indicated by the blue lines in Figure \ref{wave}.
The refraction and the reflection imply that the QFP wave and this primary EUV
wave both have wave nature. A minute later (at about 06:46 UT), the propagation
direction of the refracted wave changed when it passed through a remote
magnetic structure (see the blue circle indicated by ``P" in panel (a3) of
Figure \ref{wave}) near the AR11171. The refracted wavefront propagated in the
southeast direction, and then gradually changed its direction to the east when
passing through the strong magnetic structure indicated by ``P". The changing
propagation direction of the wave suggests the refraction effect when passing
through inhomogeneous coronal structures \citep{shen13}. The refracted
wavefront is indicated by the red lines and observed in the AIA 171 and 193
\AA\ images in Figure \ref{wave}.

\begin{figure}
\epsscale{1.0} \plotone{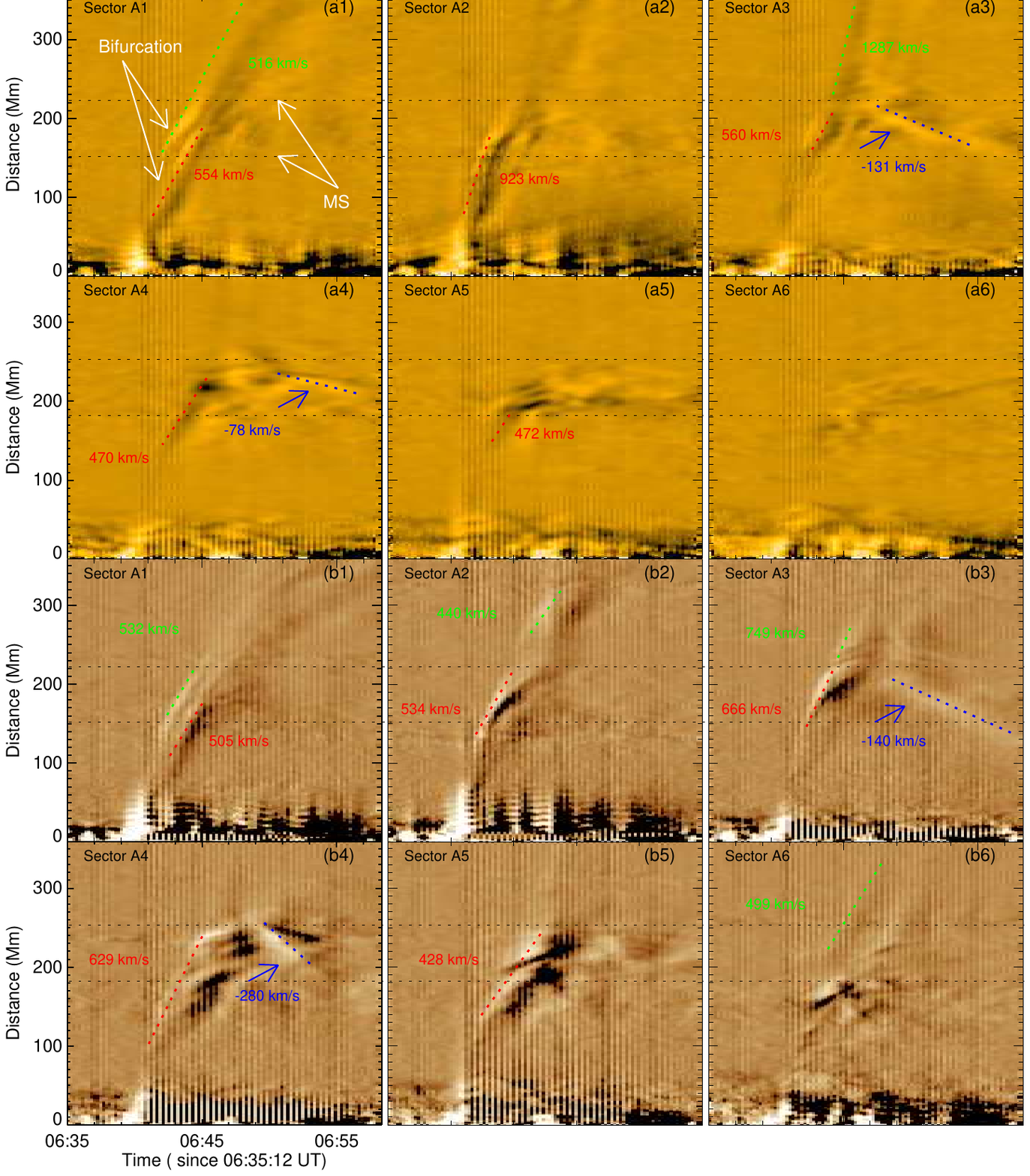}
\caption{Panels (a1)-(a6) and (b1)-(b6) are time-distance diagrams obtained from AIA 171 and 193 \AA\ running-difference
images along sectors A1-A6, respectively. Between the two black dotted lines counterpart the topological magnetic separatrix
(MS) surface structure. The speeds of the primary, refracted and reflected waves are present with different
colors. The red dotted lines are the linear fit to the bright ridge produced before the interaction of the wave and the ``MS''.
The green dotted lines are the linear fit to the bright ridge produced after the interaction of the wave and the ``MS''. The
blue dotted lines are the linear fit to the bright ridge of the reflected wave.
\label{slice1}}
\end{figure}

We use a semi-automatic method to study the detailed kinematics of the EUV
waves, as well as the reflected and refracted waves. The details of the method
can be seen in \citet{liuw10,long11,shen13}. Six sectors (A1--A6) are used to
measure the behaviors of outgoing waves and reflected wave, and another five
sectors (B1--B5) are used to measure the propagation of the refracted wave (see
panel (a) of Figure \ref{slice_position}). The results are presented in Figures
\ref{slice1} and  \ref{slice2}. Figure \ref{slice1} displays the time-distance
diagrams of the AIA 171 \AA\ and 193 \AA\ running difference intensity along
sectors A1-A6, where the two black dotted lines border the magnetic separatrix
structure. The multiple parallel wave fronts in Figures \ref{slice1}(a1) and
\ref{slice1}(b2) were also reported by \citet{shen19} and can be discerned in
the numerical simulations of \citet{chenpf2002}. The propagation speeds of
the primary wave along various sectors are in the ranges of \speed{470--923}
at 171 \AA\ (see the red dotted lines in panels (a1)-(a5)) and
\speed{428--666} at 193 \AA\ (see the black dotted lines in panels(b1)-(b6)).
The speed of the refracted wave is about \speed{1287} along sector A3, as
revealed in Figure \ref{slice1}(a3). The apparent acceleration of the
refracted wave could be caused by the change of the angle between the local
wave vector and the sector direction. The speeds of the reflected waves,
however, are only \speed{78--131} at in the \AA\ channel and \speed{140--280}
in the 193 \AA\ channel, as indicated by the blue dotted lines in Figure
\ref{slice1}. The strong reflection is due to the large gradient of the wave
speeds \citep{schmidt10,liuw12,shen12}. We also measure the propagation speeds
of the refracted wave along sectors B1--B5, and the time-distance diagrams are
displayed in Figure \ref{slice2}. One significant difference between  Figures
\ref{slice2} and \ref{slice1} is that two types of EUV waves can be discerned
along slices B1--B5 in the AIA 171 \AA\ waveband. The propagation speeds of
the fast-component EUV waves are in the ranges of \speed{412--742}, whereas
the propagation speed of the slow-component EUV waves are in the range of
\speed{229--385}. Interestingly, only the slow-component EUV waves can be
discernable in the 193 \AA\ channel, with a travelling speed of
\speed{246--390} in Figure \ref{slice2}(b1--b4). The wave signal along sector
B5 is very weak at 193 \AA.

\begin{figure}
\epsscale{1.0} \plotone{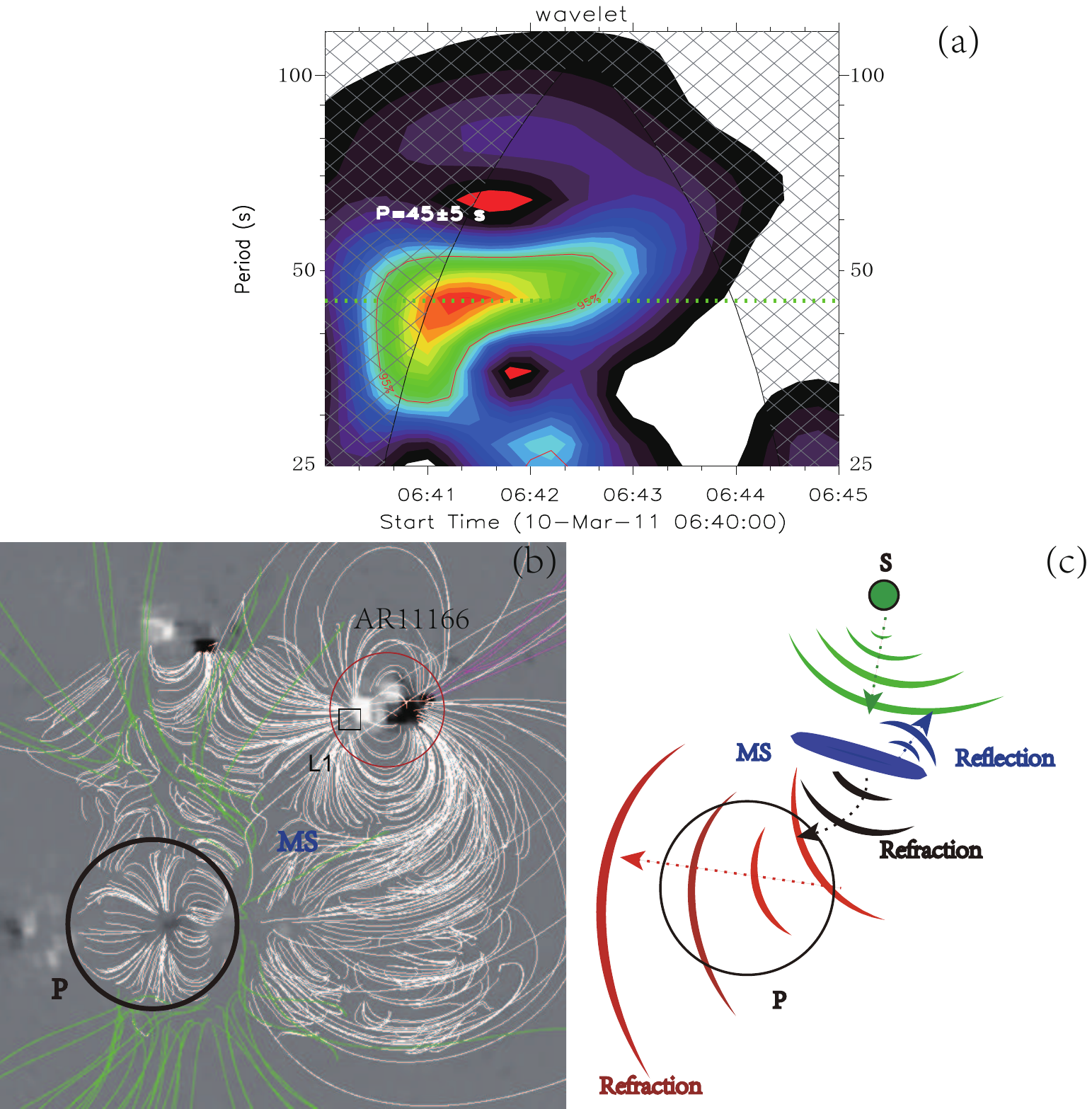}
\caption{Panel (a) shows the period of the QFP wave.
Panel (b) shows the coronal magnetic fields by using the PFSS model. The green lines
represent open field and the white lines represent closed field. The black box shows the active region of initial
flare eruption. Panel (c) presents a cartoon to further understand the evolutions of the EUV wave, the
reflection and the refraction. The ``S'' represents the source of the wave. The magnetic structure ``P'' represents the
remote magnetic structure that is enclosed by a black circle. The ``MS'' represents the topological magnetic separatrix surface.
\label{cartoon}}
\end{figure}

The kinematics of the QFP wavefronts is shown in panel (a6) of Figure
\ref{slice2} using the time-distance diagrams along the path of cut C marked
in panel (a) of Figure \ref{slice_position}. The diagram is made from the AIA
171 \AA\ running difference images. The wave train started at about 06:40:38
UT. Four wave fronts can be identified from panel (a6) of Figure \ref{slice2}.
The propagation speeds of the four wave fronts are obtained by fitting the
corresponding ridges with straight lines. The speeds of the four wavefronts
are 837, 750, 807, and 682\speed{}, respectively. The lifetime of the QFP wave
train was about 2 minutes. It was launched later than the flare, which was the
same as previous reported QFP wave trains \citep[e.g.,][]{liuw11, shen12a,
yuanding2013, kumar2017}. One important reason for many studies to support
that QFP waves are strongly associated with flares is that the QFP waves have
the same period as the flare kernels \citep{liuw11, liuw12, shen12a, shen18a,
shen18d}. In order to obtain the period of the QFP wave train, we apply the
wavelet analysis technique to the detrended intensity profile along the red
dashed line (at a distance of 150 Mm) in Figure \ref{slice2}(a6). As revealed
by Figure \ref{cartoon}(a), the period of the QFP wave train is about 45$\pm$5
seconds.

In order to better interpret the QFP waves and the EUV wave evolutions, an HMI
longitudinal magnetogram overlaid with the coronal magnetic field lines
extrapolated via the potential field source surface
\citep[PFSS][]{schatten1969, schrijver03} model is displayed in Figure
\ref{cartoon}(b), where the green (white) lines represent the open (closed)
fields. We present a cartoon to show the EUV wave propagation with the effects
of reflection and refraction in panel (c) of Figure \ref{cartoon}.

\section{Discussions and conclusions}\label{dis}

QFP waves and EUV waves are both interesting phenomena observed in the solar
corona. Generally, people considered that QFP waves have an intimate
relationship with the accompanying flare \citep{liuw11, liuw12, shen12b,
yuanding2013, kumar2017, shen18a, shen18b}. This viewpoint was supported by
the observational facts that the QFP waves have the same period as the
intensity variation of the flare kernel and the wave fronts often propagate
behind the CME frontal loop. On the contrary, although EUV waves also emanate
from flaring active regions \citep{liuw12, shen12b, kumar2013, nistico2014,
shen18a}, they, no matter the fast component or the slow component, are more
tightly associated with CMEs \citep{chenpf2002, patsourakos09, chenpf2011,
chenpf2005, zheng16a} or other types of mass eruptions, such as jets and
expanding coronal loops \citep{shen18b,shen18d}.
Taking advantage of the high spatiotemporal resolution observations from
SDO/AIA, we performed an observational study of a QFP wave train and an EUV
wave that were simultaneously excited in one flare/CME eruption event that
occurred on 2011 March 10 in the active region AR11166. The co-existence of
the two types of waves provides an excellent opportunity to investigage the
trigger mechanism of the QFP waves and EUV waves.

\begin{figure}
\epsscale{1.0} \plotone{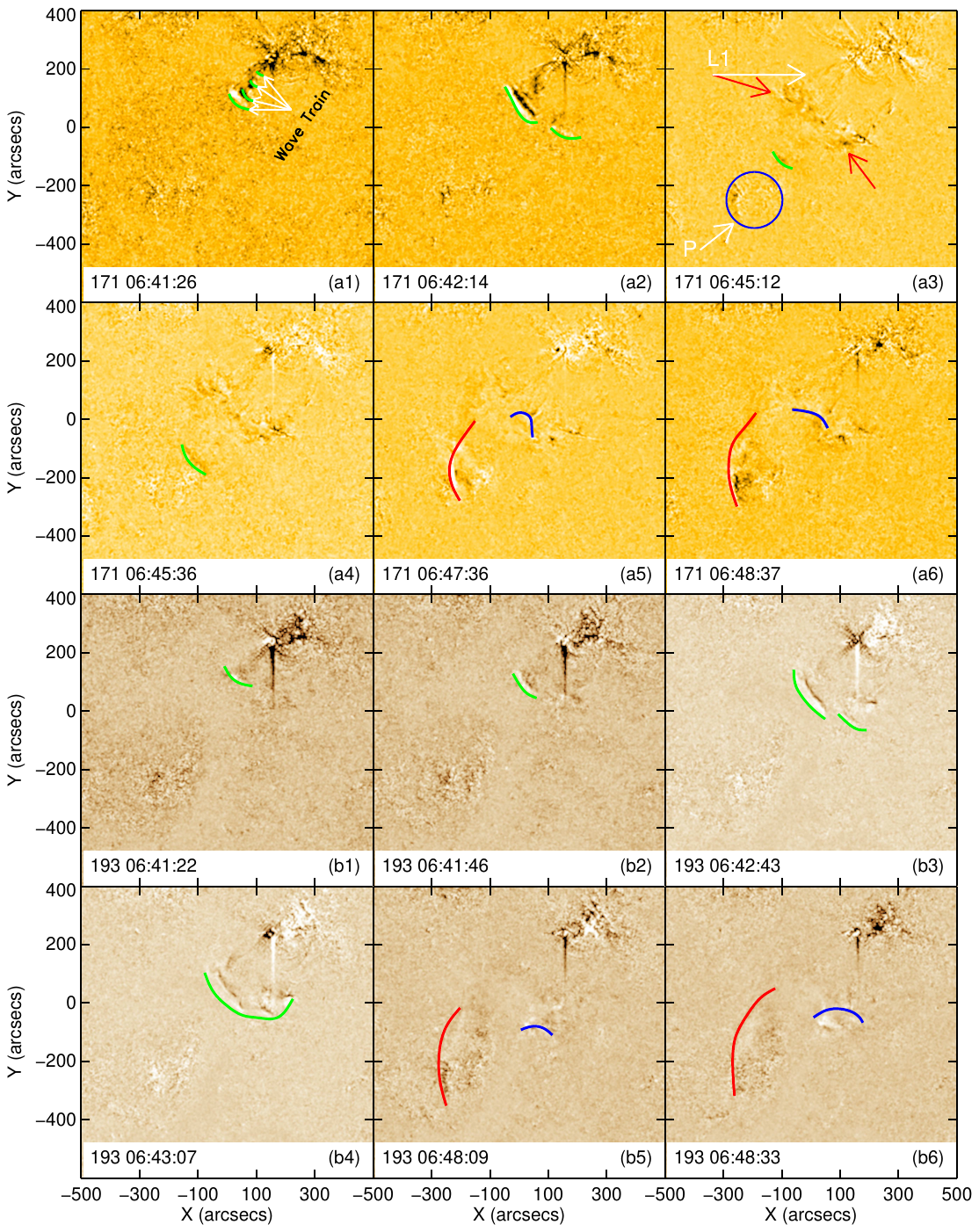}
\caption{{\sl SDO}/AIA 171 and 193 \AA\ running difference images to show the QFP and the EUV wave evolutions. Four green lines
represent the wavefronts of the QFP wave in panel (a1). The green lines display the EUV wave in panels (a2), (b1),
(b2), (b3) and (b4). The green line represents the first refracted wave in panels (a3) and (a4). The blue circle indicates the
magnetic structure (``P'') near the AR11171. The two arrows indicate the topological magnetic separatrix surface in panel (a3).
The blue and red lines represent the reflected and the second refracted waves, respectively. An animation (animation.mpeg) can
be seen in the online journal.
\label{wave}}
\end{figure}

The QFP waves were propagating along a funnel-like loop system, with a period
of about 45$\pm$5 seconds. The speeds of the QFP waves were in the range of
\speed{682--837}. According to the online animation, we can see that the QFP
wave train was only observed in the 171 \AA\ band, and was hardly visible in
the 193 \AA\ band. In contrast, the EUV waves can be identified in different
wavebands of the {\sl SDO}/AIA observations. Along the sectors A1--A6 starting
from the eruption source region, it seems that only the fast component of EUV
waves was visible, with a travelling speed of 470--923 km s$^{-1}$ in the 171
\AA\ waveband. According to \citet{chenpf2016}, such a speed range is typical
for the fast component of EUV waves. Hence it corresponds to the piston-driven
shock wave ahead of the associated CME. Combining the animation and panels
(a1) and (b1) of Figure \ref{wave}, we believe that this EUV wavefront was
in juxtaposition with the first wavefront of the QFP wave train. This implies
that the fast-component of the EUV waves and the QFP wave train in this event
originate from the same process, i.e., they are driven by the associated CME,
rather than the pressure pulse in the associated flare kernels. Moreover, the
period of the QFP wave train was about 45$\pm$5 s, and such a period is absent
in the light curve of the flare kernel. Therefore, we propose that CME-driven
shock wave, when interacting with funnel-like coronal loops, may also drive
QFP wave trains. In this scenario, the period of the QFP wave train is not
associated with the fluctuating brightness of the flare kernels.

\begin{figure}
\epsscale{1.0}
\plotone{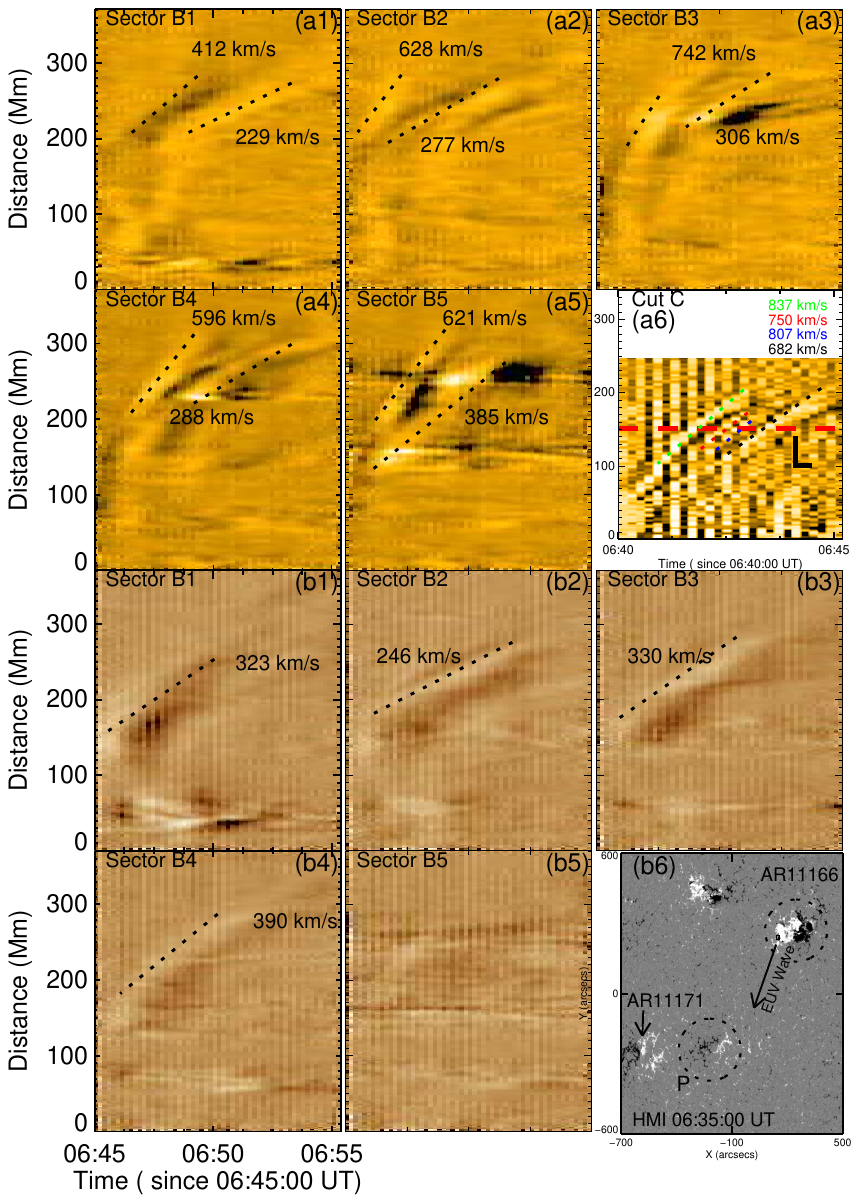}
\caption{Panels (a1)-(a5) and (b1)-(b5) are time-distance diagrams obtained from AIA 171 and 193 \AA\ running difference
images along sectors B1-B5, respectively. Panel (a6) shows the time-distance diagram obtained from AIA 171 \AA\
running-difference images along Cut C. The red-dashed line (L) in panel (a6) mark the position where we analyze the
periodicity of the QFP wave. Panel (b6) simply shows magnetic configurations of AR11166 and the magnetic structure ``P''. The
red arrow indicates the direction of the initial wave propagates direction. The magnetic structure ``P'' represents the magnetic
structure near the AR11171.
 \label{slice2}}
\end{figure}

Besides, the effects of refraction and reflection are observed when the
fast-component of the EUV wave interacted with a magnetic separatrix. The
speeds of primary, reflected, and refracted waves are 428--923 km s$^{-1}$,
78--280 km s$^{-1}$, and 246--742 km s$^{-1}$, respectively. Across the
magnetic separatrix, the propagation direction of the wavefront gradually
changed from southeastward to eastward. \citet{uchida1968} and
\citet{afanasyev2011} indicated that if a wave is a true fast-mode
magnetosonic wave, refraction would veer the wave from regions with higher
wave speeds toward regions with lower wave speeds. All above results are
suggestive of that the EUV waves along sectors A1--A6 are true fast
magnetosonic waves. According to the prediction of \citet{chenpf2002, chenpf2005},
two types of EUV waves would be associated with a CME event, which was later
confirmed by {\sl SDO}/AIA observations \citep[e.g.,][]{chenpf2011, asai12,
cheng12, kumar2013}. However, it should be noted that the two components of EUV
waves do not always appear simultaneously, and only one component is evident
in some events. In this paper, it seems that only the fast-component EUV wave
appeared along sectors A1--A6. However, as revealed by Figure
\ref{slice2}(a1--a5), two components of EUV waves can be clearly identified
in the 171 \AA\ images, with the fast-component travelling with a speed of
412--742 km s$^{-1}$, and the slow-component travelling with a speed of
\speed{229--385}  km s$^{-1}$. The speed of the slow-component wave is roughly three
times smaller than that of the fast-component EUV wave, as predicted by the
magnetic field-line stretching model predicted by \citet{chenpf2002,
chenpf2005}.

In summary, QFP wave train and EUV waves in a single event was analyzed in this paper.
It was revealed that the EUV wavefront was in juxtaposition with the first wavefront
of the QFP wave train. \textbf{Hence, our finding confirms that QPF waves can be excited by CMEs.}
During the interaction, the initial broadband pulse can dispersively evolve into multiple QFP wavefronts.
\citet{pascoe2013} presented simulation work that showed this process, and \citet{nistico2014} further
confirmed the scenario via observation. Here, the broadband pulse could be regarded as a piston-driven
shock wave as observed in the present case, which further dispersively evolved into the QFP wave along
the funnel-like coronal loops. This scenario is different from that proposed by \citet{liuw11} in that
the period of the QFP wave train has nothing to do with the flare kernel. More events are required in
order to understand what determines the period of the QFP wave train in this scenario.

\acknowledgments We thank the excellent data provided by the {\em SDO} and {\em STEREO} teams.
 We also thank the referee for his/her valuable suggestions and comments that improved the quality of the letter.
 This work is funded by the grants from the National Scientific Foundation of China (NSFC 11533009)
and from the Key Laboratory of Geospace Environment, CAS, University of Science and Technology of China. This work is also
supported by the grant associated with project of the Group for Innovation of Yunnan Province and the Strategic Priority
Research Program of CAS with grant XDA-17040507. The authors Y. Liu and Z.Z. Abidin would like to thank the University of
Malaya Faculty of Science grant (GPF040B-2018) for their support. The research by A. Elmhamdi was supported by King Saud
University, Deanship of Scientific Research, College of Science Research Center. In addition, we are also grateful to the
One Belt and One Road project of the West Light Foundation, CAS.



\end{document}